# Distributed Rateless Codes with UEP Property

Ali Talari and Nazanin Rahnavard

Oklahoma State University, Stillwater, OK 74078

Emails: {ali.talari, nazanin.rahnavard}@okstate.edu

## Abstract

When multiple sources of data need to transmit their *rateless coded* symbols through a single relay to a common destination, a *distributed rateless code* instead of several separate conventional rateless codes can be employed to encode the input symbols to increase the transmission efficiency and flexibility.

In this paper, we propose *distributed rateless codes (DU-rateless)* that can provide *unequal error protection* (UEP) for distributed sources with different data block lengths and different importance levels. We analyze our proposed DU-rateless codes employing *And-Or tree analysis* technique. Next, we design several sets of optimum DU-rateless codes for various setups employing multi-objective genetic algorithms and evaluate their performances.

## I. Introduction

*Rateless codes* [1], [2], [3] are modern and efficient *forward error correction* (FEC) codes. Each rateless code is determined by a *degree distribution*, which is precisely designed to achieve a capacity-approaching performance.

In distributed data transmission using rateless codes, $r$ data sources need to transmit their rateless encoded symbols to a destination through a common relay. In general, $r$ sources may have different data block lengths and different data importance levels, which necessitate the design of flexible *distributed rateless codes* that can provide *unequal error protection* (UEP) of data for different sources. In this paper, we propose *distributed UEP rateless codes* (DU-rateless codes), which are a realization of such codes.

It has been shown that the efficiency of rateless codes increases as the data block length increases [1], [2], [3], [4]. Thus, in distributed rateless codes it is advantageous to *combine* the incoming symbols in the intermediate relay, which is equivalent to coding a larger data block. Moreover, by tuning coding parameters in each data source and parameters of the relay, UEP property can be provided for different data sources. The problem in DU-rateless codes is to optimally design different degree distributions for each source and to design relaying parameters to realize the desired UEP property and a minimal error rate for all data sources.

Previously, two contributions [4] and [5] have studied distributed rateless codes. Authors in [4] have designed *distributed LT codes*. In the proposed scheme in [4], the relay combines all incoming symbols that are coded at $r \in \{2, 4\}$ sources with the same degree distribution. This coding degree distribution is designed such that the degree distribution of the combined symbols at the relay follows an optimum degree distribution called *Robust-Soliton* degree distribution [1]. This algorithm cannot provide UEP for different sources and obligates sources to have the same data block lengths.

In [5], authors have also considered the case where the source nodes have the same data lengths, and all source nodes perform the encoding with the same degree distribution. Authors have also studied the case where the relay generates final output symbols with another independent degree distribution that determines how many symbols should be combined in the relay to generate an output symbol. The optimization for relaying and coding parameters has been performed separately in this paper, which may result in suboptimal performance.

In this paper, we take several steps further compared to [4], [5], and propose DU-rateless codes that are inspired by UEP rateless codes [6], [7]. DU-rateless codes are able to provide UEP for different sources that may also have various data block lengths.

The paper is organized as follows. In Section II, we propose DU-rateless codes and analyze these codes employing *And-Or tree analysis* technique [8]. In Section III, we design and evaluate the performance of several ensembles of DU-rateless codes for different UEP setups by optimizing degree distributions for each source and relaying parameters along each other employing the state-of-the-art *multi-objective genetic algorithms* NSGA-II [9]. Finally, Section IV concludes the paper.

## II. DU-RATELESS CODES

Rateless codes can generate a limitless number of output symbols from $k$ input symbols based on a degree distribution $\{\Omega_1, \Omega_2, \ldots, \Omega_k\}$, where $\Omega_i$ is the probability that an output symbol has degree $i$, and $\sum_{i=1}^{k} \Omega_i = 1$. This probability distribution can also be shown by its generator polynomial $\Omega(x) = \sum_{i=1}^{k} \Omega_i x^i$. In rateless coding, first an output symbol degree $d$ is randomly chosen from $\Omega(x)$. Next, $d$ input symbols are chosen uniformly at random from $k$ input symbols and are *XOR*ed together to generate an output encoded symbol. $\Omega(x)$ is usually finely tuned such that $k$ input symbols can be decoded from any $\gamma k$ collected output symbols at decoder, where $\gamma$ is a number slightly larger than one and is called *coding overhead*.

Rateless decoding process consists of one step: Find an output symbol such that the value of all but one of its neighboring input symbols is known. The value of the unknown input symbol is computed by a simple XOR. We apply this step until no more such output symbols can be found.

In DU-rateless coding, each source performs rateless coding with a distinct degree distribution on its data block and forwards its output symbols to the relay. For the sake of simplicity in analytical expressions, we consider a case with $r = 2$. Consider a distributed data transmission with two sources $s_1$ and $s_2$, and data block lengths $\rho k$ and $k$, respectively, where $0 < \rho \leq 1$. $s_1$ and $s_2$ encode their input symbols with degree distributions $\Omega(x)$ and $\varphi(x)$ with the largest degrees $B_1$ and $B_2$, respectively, and forward them to the relay (see Figure 1). Relay $R$ receives output symbols from two sources and performs as follows.

1) With probabilities $p_1$ and $p_2$ it relays the first and the second source's output symbol to the destination D, respectively.

2) With probability $p_3 = 1 - p_1 - p_2$ it combines two incoming symbols and forwards the combined symbol to the destination.

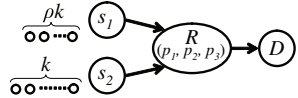

Fig. 1. Adopted model for DU-rateless codes. $s_1$, $s_2$, $R$, and $D$ represent distributed sources, relay, and destination, respectively.

The proposed DU-rateless code ensemble is specified by parameters $(\rho k, k, \Omega(x), \varphi(x), p_1, p_2, p_3, \gamma)$. DU-rateless decoding is the same as rateless decoding. The decoding is successful when $(1+\rho)\gamma k$ output symbols are received at the destination.

Following [3], we may view the input and output symbols as vertices of a bipartite graph $G$, where the input symbols are the variable nodes and the output symbols are the check nodes. Without loss of generality, throughout this paper we may assume that the symbols are binary symbols for simplicity.

In DU-rateless coding described above, the corresponding bipartite graph at the receiver has two types of variable nodes (input symbols from $s_1$ and $s_2$), and three types of check nodes generated by the relay as depicted in Figure 1. The check nodes in the first group are generated based on $\Omega(x)$ and are only connected to input symbols of $s_1$. Similarly, the check nodes in the second group are generated based on $\varphi(x)$ and are only connected to input symbols of $s_2$. Finally, the check nodes in the third group are generated using input symbols from both $s_1$ and $s_2$ with a degree distribution equal to $\Omega(x) \times \varphi(x)$ [4]. It is worth noting that a check node belongs to the first, second, and third group with probabilities $p_1, p_2$, and $p_3$, respectively.

To investigate the recovery probability of an input symbol in DU-rateless codes, we first extend And-Or tree analysis [8], [3], [6] technique to fit our problem. Then, we map decoding of DU-rateless codes to extended And-Or tree analysis, and evaluate the recovery probability of input symbols.

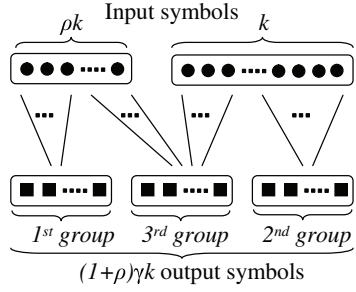

Fig. 2.   The bi-partite graph representing input and output symbols for $r = 2$.

### A. And-Or Tree Analysis Technique

Consider two And-Or trees [8] $T_{l,1}$ and $T_{l,2}$ with depth $2l$. Assume that $T_{l,1}$ and $T_{l,2}$ have Type-X and Type-Y OR-nodes and Type-I, Type-II, and Type-III AND-nodes. For each tree, the root of the tree is at depth $0$, its children are at depth $1$, their children at depth $2$, and so forth. Each node at depth $0, 2, 4, \ldots, 2l-2$ is an *OR-node* (and it evaluates logical OR operation on the value of its children), and each node at depth $1, 3, 5, \ldots, 2l-1$ is called an *AND-node* (and it evaluates logical AND operation on the value of its children). The root of $T_{l,1}$ is a Type-X OR-node, and the root of $T_{l,2}$ is a Type-Y OR-node as depicted in Figures 3 and 4, respectively.

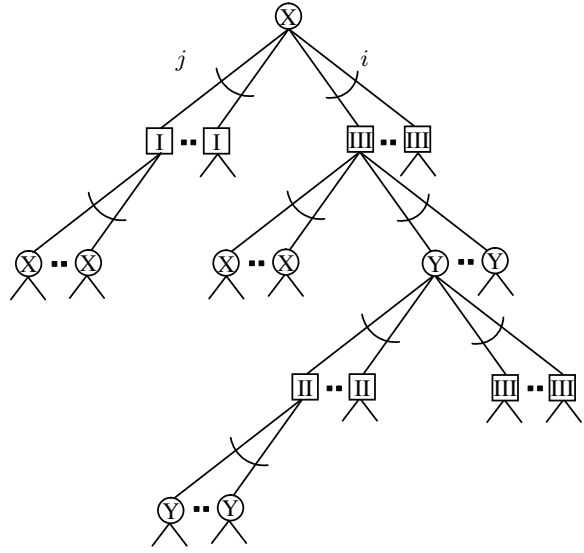

PSfrag replacements

Fig. 3.   $T_{l,1}$ And-Or tree with two types of OR-nodes and three types of AND-nodes with a Type-X OR-node root.

We assume that in both $T_{l,1}$ and $T_{l,2}$, Type-X OR-nodes choose $i \in \{0, \ldots, A_1\}$ and $j \in \{0, \ldots, A_1\}$ children from Type-I and Type-III AND-nodes with probabilities $\delta_{i,1}$ and $\delta_{j,1}$, respectively. Furthermore, Type-Y OR-nodes choose $i \in \{0, \ldots, A_2\}$ and $j \in \{0, \ldots, A_2\}$ children from Type-II and Type-III AND-nodes with probabilities $\delta_{i,2}$ and $\delta_{j,2}$, respectively.

Further, Type-I AND-nodes choose $i \in \{0, \ldots, B_1-1\}$ children from Type-X OR-nodes with probability $\beta_{i,1}$, and Type-II AND-nodes choose $i \in \{0, \ldots, B_2-1\}$ children from Type-Y OR-nodes with probability $\beta_{i,2}$.

Moreover, in $T_{l,1}$, Type-III AND-nodes choose $j \in \{0, \ldots, B_1-1\}$ and $i \in \{1, \ldots, B_2\}$ children from Type-X and Type-Y OR-nodes with probabilities $\beta_{j,1}$ and $\beta_{i,3}$, respectively. Note that Type-III AND-nodes in $T_{l,1}$ should have at least one child from Type-Y OR-nodes, since otherwise it is a Type-I AND-node. In addition, in $T_{l,2}$, Type-III AND-nodes can choose $j \in \{0, \ldots, B_2-1\}$ and $i \in \{1, \ldots, B_1\}$ children from Type-Y and Type-X OR-nodes with probabilities $\beta_{j,2}$ and $\beta_{i,4}$, respectively. Similar to Type-III AND-nodes in $T_{l,1}$, Type-III AND-nodes in $T_{l,2}$ need to have at least one child from Type-X OR-nodes to be distinguished from Type-II AND-nodes.

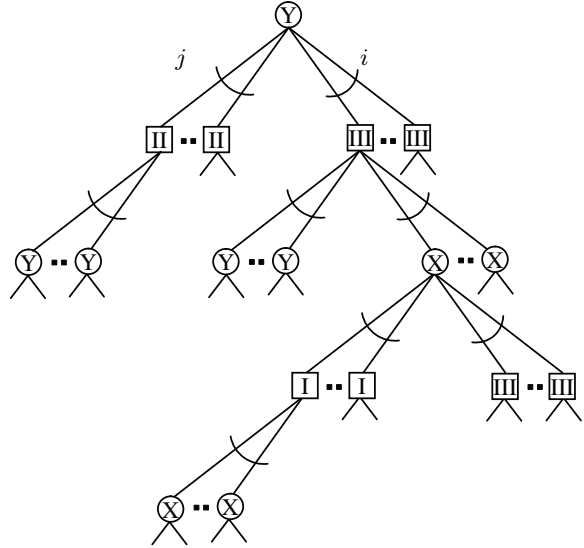

Fig. 4. $T_{l,2}$ And-Or tree with two types of OR-nodes and three types of AND-nodes with a Type-Y OR-node root.

Finally, we assume that in both $T_{l,1}$ and $T_{l,2}$ the ratio of the number of AND-nodes of Type-I, Type-II, and Type-III is $p_1$, $p_2$, and $p_3 = 1 - p_1 - p_2$, where $0 \leq p_i \leq 1, \forall i$.

Type-X and Type-Y OR-nodes at depth $2l$ are independently assigned a value of $0$ with probabilities $y_{0,1}$ and $y_{0,2}$, respectively. Also OR-nodes with no children are assumed to have a value $0$, whereas AND-nodes with no children are assumed to have a value $1$. We are interested in finding $y_{l,1}$ and $y_{l,2}$, the probabilities that the root nodes of $T_{l,1}$ and $T_{l,2}$ evaluate to $0$, respectively, if we treat the trees as a Boolean circuits. Lemma 1 formulates $y_{l,1}$ and $y_{l,2}$.

*Lemma 1:* Let $y_{l,1}$ and $y_{l,2}$ be the probabilities that the roots of And-Or trees $T_{l,1}$ and $T_{l,2}$ evaluate to $0$, respectively. Then

$$y_{l,1} = \delta_1 \left( 1 - p_1' \sum_{i=0}^{B_1-1} \beta_{i,1}(1 - y_{l-1,1})^i - \right.$$
$$\left. p_3' \sum_{d=0}^{B_1+B_2-2} \sum_{j=0}^{d} [\beta_{j,1}(1 - y_{l-1,1})^j \beta_{d-j+1,3}(1 - y_{l-1,2})^{d-j+1}] \right),$$
$$y_{l,2} = \delta_2 \left( 1 - p_2' \sum_{i=0}^{B_2-1} \beta_{i,2}(1 - y_{l-1,2})^i - \right.$$
$$\left. p_4' \sum_{d=0}^{B_1+B_2-2} \sum_{j=0}^{d} [\beta_{j,2}(1 - y_{l-1,2})^j \beta_{d-j+1,4}(1 - y_{l-1,1})^{d-j+1}] \right),$$

(1)

with $\delta_1(x) = \sum_{i=0}^{A_1} \delta_{i,1} x^i$, $\delta_2(x) = \sum_{i=0}^{A_2} \delta_{i,2} x^i$, $p_1' = \frac{p_1}{1-p_2}$, $p_3' = \frac{1-p_1-p_2}{1-p_2} = \frac{p_3}{1-p_2}$, $p_2' = \frac{p_2}{1-p_1}$ and $p_4' = \frac{1-p_1-p_2}{1-p_1} = \frac{p_3}{1-p_1}$.

*Proof:* The proof is straight forward and similar to the proof of [7, Lemma 2], and is not included in this paper due to space limit. ∎

The relation between the above analysis and the error probabilities for DU-rateless codes is given in the following.

### B. Analysis of DU-rateless Codes

In this section, we examine the DU-rateless codes under iterative decoding. Let $G$ denote the bipartite graph corresponding to a DU-rateless code at the receiver. In [2], [3], [7], [6], authors have shown that iterative belief propagation decoding of rateless codes can be rephrased as following. At every iteration of the algorithm, messages (0 or 1) are sent along the edges from check nodes to variable nodes, and then from variable nodes to check nodes.

A variable node sends 0 to an adjacent check node if and only if its value is not recovered yet. Similarly, a check node sends 0 to an adjacent variable node if and only if it is not able to recover the value of the variable node. In other words, a variable node sends 1 to a neighboring check node if only if it has received at least one message with value 1 from its other neighboring check nodes. Also a check node sends 0 to a neighboring variable node if only if it has received at least one message with value 0 from its other neighboring variable nodes. Therefore, we see that variable nodes indeed do the logical OR operation, and the check nodes do the logical AND operation.

Consequently, we can use the results of Lemma 1 on a subgraph $G_{l,1}$ of $G$ to find the probability that a $s_1$ variable node is not recovered after $l$ decoding iterations (its value evaluates to zero). We choose $G_{l,1}$ as following. Choose an edge $(v, w)$ uniformly at random from all edges in $G$ with one end among variable nodes of $s_1$. Call the variable node $v$ the root of $G_{l,1}$. Subgraph $G_{l,1}$ is the graph induced by $v$ and all neighbors of $v$ within distance $2l$ after removing the edge $(v, w)$. It can be shown that $G_{l,1}$ is a tree asymptotically [8]. We can map encoded symbols from $s_1$, encoded symbols from $s_2$, and combined encoded symbols in $G_l$ to Type-I, Type-II, and Type-III AND-nodes in $T_{l,1}$, respectively. Further, variable nodes of $s_1$ and $s_2$ in $G_{l,1}$ can be mapped to Type-X and Type-Y OR-nodes in $T_{l,1}$, respectively.

In the same way, to find the probability that a $s_2$ variable node is not recovered after $l$ decoding iterations, we choose a subgraph $G_{l,2}$ of $G$ similar to $G_{l,1}$ except that we choose the edge $(v, w)$ such that it has an end among variable nodes of $s_2$. $G_{l,2}$ can be mapped to $T_{l,2}$ in the same way that $G_{l,1}$ is mapped to $T_{l,1}$.

To complete DU-rateless codes analysis, we only need to compute the probabilities $\beta_{i,1}$, $\beta_{i,2}$, $\beta_{i,3}$, $\beta_{i,4}$, and functions $\delta_1(x)$ and $\delta_2(x)$, which are given in the following Lemma.

*Lemma 2:* Consider trees $T_{l,1}$ and $T_{l,2}$ that are derived based on a $(\rho k, k, \Omega(x), \varphi(x), p_1, p_2, p_3, \gamma)$ DU-rateless code graph $G$. The probabilities $\beta_{i,1}$, $\beta_{i,2}$, $\beta_{i,3}$, $\beta_{i,4}$, and functions $\delta_1(x)$ and $\delta_2(x)$ are given as

$$\delta_1(x) = e^{(1-p_2)\mu_1\gamma\frac{(1+\rho)}{\rho}(x-1)}, \delta_2(x) = e^{(1-p_1)\mu_2\gamma(1+\rho)(x-1)},$$

$$\beta_{i,1} = \frac{(i+1)\Omega_{i+1}}{\Omega'(1)}, \beta_{i,2} = \frac{(i+1)\varphi_{i+1}}{\varphi'(1)},$$

$$\beta_{i,3} = \varphi_i, \text{ and } \beta_{i,4} = \Omega_i,$$

where $\mu_1 = \Omega'(1)$ and $\mu_2 = \varphi'(1)$ are the average degrees of the two coding degree distributions $\Omega(x)$ and $\varphi(x)$.

*Proof:* We have $\beta_{i,1}$ is the probability that a randomly chosen edge in $T_{l,1}$ is connected to a Type-I or a Type-III AND-node with $i$ children among Type-X OR-nodes. This is the probability that the edge is connected to a Type-I or Type-III AND-node of degree $i + 1$ (excluding edges connected to Type-Y OR-nodes from Type-III AND-nodes). It can be seen that out of $\gamma\rho k\Omega'(1)$ total edges connected to Type-I and Type-III AND-nodes from Type-X OR-nodes, $\gamma\rho k(i+1)\Omega_{i+1}$ edges are connected to AND-nodes of degree $i + 1$. Therefore, we have $\beta_{i,1} = \frac{(i+1)\Omega_{i+1}}{\Omega'(1)}$. Similarly, we have $\beta_{i,2} = \frac{(i+1)\varphi_{i+1}}{\varphi'(1)}$. Moreover, $\beta_{i,3}$ is the probability the a randomly chosen edge in $T_{l,1}$ is connected to a Type-III AND-node with $i$ children in the Type-Y OR-node. This simply gives $\beta_{i,3} = \varphi_i$. In the same way, $\beta_{i,4} = \Omega_i$.

We have $\delta_{i,1}$ is the probability that the variable node connected to a randomly selected edge has degree $i + 1$ given that the variable node belongs to Type-X OR-nodes. The total number of edges connected to Type-X OR-nodes is $\mu_1\gamma k(1-p_2)\frac{(1+\rho)}{\rho}$ out of which $(i+1)\lambda_{i+1,1}k$ edges are connected to OR-nodes of degree $i + 1$, where $\lambda_{i+1,1}$ is the probability that a variable node of $s_1$ has degree $i + 1$. We observe that $\lambda_{d,1} = \binom{(1-p_2)\mu_1\gamma k(1+\rho)}{d}(\frac{1}{\rho k})^d(1-\frac{1}{\rho k})^{(1-p_2)\mu_1\gamma k(1+\rho)-d}$ since $\mu_1(1+\rho)\gamma k(1-p_2)$ edges are connected uniformly at random to $s_1$'s variable nodes. Therefore, we have $\delta_{i,1} = \frac{(i+1)\lambda_{i+1,1}}{\mu_1\gamma(1-p_2)\frac{(1+\rho)}{\rho}}$. After substitution, we have $\delta_1(x) = e^{(1-p_2)\mu_1\gamma\frac{(1+\rho)}{\rho}(x-1)}$. Similarly, we have $\delta_2(x) = e^{(1-p_1)\mu_2\gamma(1+\rho)(x-1)}$. ∎

Lemma 1 and Lemma 2 give two sequences $\{y_{l,1}\}_l$ and $\{y_{l,2}\}_l$, which are decreasing and convergent with respect to the number of decoding iterations, $l$ [7], [6]. Let $BER_1$ and $BER_2$ denote the corresponding fixed points. These fixed points are the probabilities that Type-X and Type-Y OR-nodes are not recovered after $l$ decoding iterations. In other words, these fixed points are the final decoding error rates of a $(\rho k, k, \Omega(x), \varphi(x), p_1, p_2, p_3, \gamma)$ DU-rateless code.

## III. DU-RATELESS CODES DESIGN

In this section, we employ our analytical results in the previous section to design *optimal DU-rateless code parameters* for different setups. For DU-rateless coding with $r = 2$, two error rates $BER_1$ and $BER_2$ are defined. The values of these two error rates are *dependant*, i.e. improving one error rate by modifying DU-rateless code parameters may result in degrading the other error rate. In other words, we are dealing with two dependant error rates. Consequently, if we consider error rates as conflicting objective functions, we have a multi-objective optimization problem.

Since we have more than one objective functions to minimize, we need to employ *pareto optimality* concept. In Figure 5, we have depicted a simple minimization problem with two *conflicting* objective functions and two decision variables. Assume that shaded area in decision space is mapped to the shaded area in objective space. We can observe that, three sets of variables shown on the decision space result in $F_1$'s and $F_2$'s that no other decision variables can concurrently surpass. These solution are called *pareto optimal* or *non-dominated* solutions, and their mapping to objective space is called *pareto front*. We can observe that in contrast to single objective optimizations, we can have infinite number of optimum decision variables.

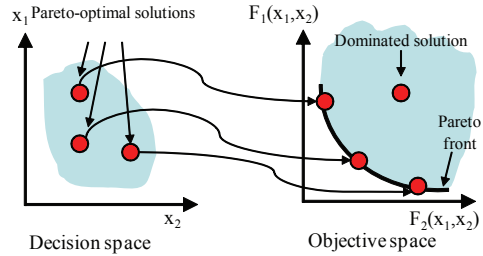

Fig. 5. Concept of pareto optimality, pareto front, and domination for a two-objective minimization problem with two decision variables, $x_1$ and $x_2$.

Multi-objective optimization methods such as *NSGA-II* [9] search to find solutions that result in pareto front.

### A. DU-rateless Codes Design Employing NSGA-II

We fix the parameters $\gamma = 1.05$ and $B_1 = B_2 = 100$, and employ the state-of-the-art multi-objective genetic algorithm NSGA-II [9] to find the optimum value for $\Omega(x)$ and $\varphi(x)$ along with relaying parameters $p_1$, $p_2$, and $p_3$ that minimize $BER_1$ and $BER_2$ for various values of $\eta = \frac{BER_2}{BER_1}$ and $\rho \in \{0.3, 0.5, 1\}$. In other words, we have a problem including two objective functions, $BER_1$ and $BER_2$, with 202 independent decision variables, i.e. $\bar{x} = \{\Omega_1, \Omega_2, \dots, \Omega_{100}, \varphi_1, \varphi_2, \dots, \varphi_{100}, p_1, p_2\}$.

The resulting pareto fronts for $\rho \in \{0.3, 1\}$ are depicted in Figure 6 in objective space.

Note that each point in Figure 6 embodies two degree distributions and three relaying parameters, and none of these points dominate another member in the pareto front. One should choose an appropriate point according to a desired $\eta$ (UEP parameter), and employ the corresponding DU-rateless code. Since optimization results cannot be reported in the paper due to huge number of members, they are made available online at [10] for $\rho \in \{0.3, 0.5, 1\}$. Note that $\eta = 1$ corresponds to *equal error protection* (EEP) case where data from $s_1$ and $s_2$ are equally protected.

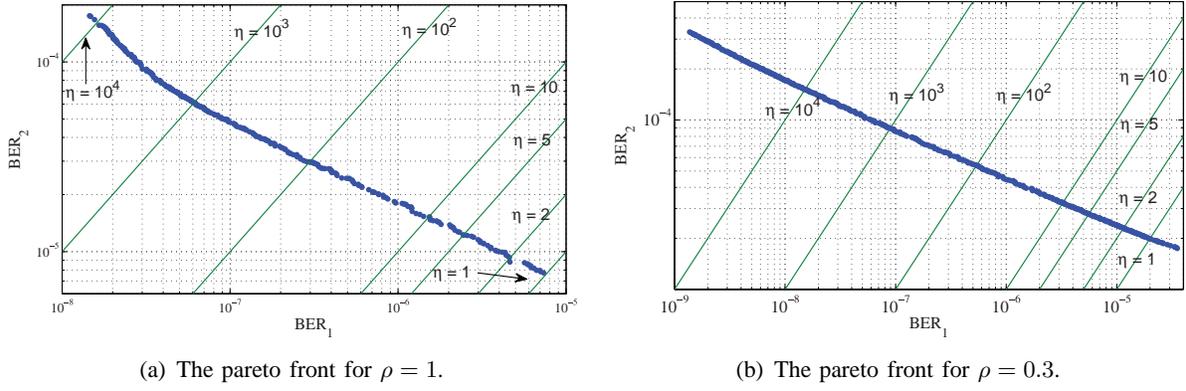

(a) The pareto front for $\rho = 1$.

(b) The pareto front for $\rho = 0.3$.

Fig. 6. The resulting pareto fronts for DU-rateless codes design. Each point represents one set of DU-rateless codes parameters $\Omega(x)$, $\varphi(x)$, and relaying parameters $p_1$, $p_2$, and $p_3$ for $\gamma = 1.05$ and $\rho \in \{0.3, 0.5, 1\}$.

## B. Performance Evaluation of Designed Codes

From the sets of optimized degree distributions available at [10], we choose DU-rateless codes for $\eta \in \{10, 100\}$ and $\rho = 1$ and evaluate their performance in Figure 7. For comparison, we have also provided $BER_1$ and $BER_2$ for EEP case ($\eta = 1$).

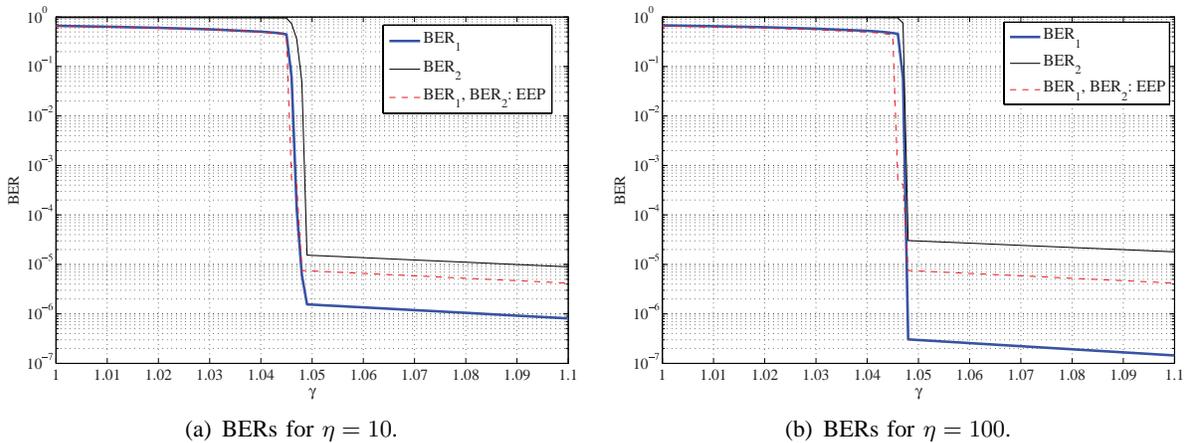

(a) BERs for $\eta = 10$.

(b) BERs for $\eta = 100$.

Fig. 7. The resulting BERs with optimized sets of parameters for $\eta \in \{10, 100\}$, $\gamma = 1.05$, and $\rho = 1$.

Figure 7 illustrates that the expected UEP property is fulfilled for $\gamma = 1.05$ with the minimal values of $BER_1$ and $BER_2$. The parameters of a DU-rateless code for $\rho = 1$ and $\eta = 10$ are given as $p_1 = 0.4822$, $p_2 = 0.1173$, $p_3 = 0.4005$,

$$\begin{aligned}
\Omega(x) = {} & 0.039x^1 + 0.492x^2 + 0.094x^3 + 0.09x^4 + 0.096x^5 \\
& + 0.002x^6 + 0.055x^7 + 0.019x^8 + 0.033x^9 + 0.014x^{10} \\
& + 0.004x^{20} + 0.006x^{27} + 0.005x^{31} + 0.005x^{43} + 0.005x^{78} \\
& + 0.005x^{86} + 0.014x^{95} + 0.007x^{100},
\end{aligned}$$

and

$$\begin{aligned}
\varphi(x) = {} & 0.072x^1 + 0.48x^2 + 0.055x^3 + 0.051x^4 + 0.063x^5 \\
& + 0.059x^6 + 0.037x^7 + 0.026x^8 + 0.025x^9 + 0.036x^{10} \\
& + 0.005x^{15} + 0.003x^{28} + 0.005x^{37} + 0.002x^{44} + 0.002x^{70} \\
& + 0.002x^{77} + 0.003x^{83} + 0.004x^{93} + 0.052x^{95} + 0.002x^{97},
\end{aligned}$$

We can see that to achieve an optimum distributed coding $40.05\%$ of generated symbols should be combined in the relay. The performance of this code is illustrated in Figure 7(a).

We emphasis that the interesting point of our approach is optimizing all codes' parameters along each other using multi-objective genetic algorithms. Note that conventional linear programming optimization methods may not be able to optimize all parameters of our code at the same time resulting in suboptimal code design.

Further, finding a general analytical expression for $r > 2$ is the next step in our future research.

## IV. Conclusion

In this paper, we proposed distributed rateless codes with *unequal error protection* (UEP) property. Besides providing UEP property, DU-rateless codes do not need that all sources to have equal data block lengths. First, we analyzed DU-rateless codes employing And-Or tree analysis technique, and then we designed several optimum sets of DU-rateless codes using multi-objective genetic algorithms. Finally, we evaluated the designed codes employing simulation results. Simulation results show that DU-rateless codes fulfill the expected UEP property with minimal error rates.